# Anomalous twin boundaries in 2D materials

*A. P. Rooney[1], Z. Li[1,2], W. Zhao[3,6], A. Gholinia[1], A. Kosikov[4,5], G. Auton[2,4], F. Ding[3,6], R. V. Gorbachev[2,5], R. J. Young[1,2,3], S. J Haigh[1,2]\**

1. School of Materials, University of Manchester, Manchester M13 9PL, UK
2. National Graphene Institute, University of Manchester, Manchester M13 9PL, UK
3. Institute of Textiles and Clothing, Hong Kong Polytechnic University, Hung Hom, Hong Kong
4. Manchester Centre for Mesoscience and Nanotechnology, University of Manchester, Manchester M13 9PL, UK
5. School of Physics and Astronomy, University of Manchester, Oxford Road, Manchester, M13 9PL, UK
6. Center for Multidimensional Carbon Materials, Institute for Basic Science (IBS-CMCM)/School of Material Science and Engineering, Ulsan National Institute of Science and Technology (UNIST), Ulsan 44919, Korea

The high mechanical strength and excellent flexibility of 2D materials such as graphene are some of their most important properties[1]. Good flexibility is key for exploiting 2D materials in many emerging technologies, such as wearable electronics, bioelectronics, protective coatings and composites[1] and recently bending has been suggested as a route to tune electronic transport behaviour[2]. For virtually all crystalline materials macroscopic deformation is accommodated by the movement of dislocations and through the formation of twinning defects[3]; it is the geometry of the resulting microstructure that largely determines the mechanical and electronic properties. Despite this, the atomic microstructure of 2D materials after mechanical deformation has not been widely investigated: only by understanding these deformed microstructures can the resulting properties be accurately predicted and controlled. In this paper we describe the different structural features that can form as a result of bending in van der Waals (vdW) crystals of 2D materials. We show that twin boundaries, an important class of crystal defect, are delocalised by several nm and not atomically sharp as has been assumed for over half a century[4]. In addition, we demonstrate that different classes of microstructure are present in the deformed material and can be predicted from just the atomic structure, bend angle, and flake thickness. We anticipate that this new knowledge of the deformation structure for 2D materials will provide foundations for tailoring transport behaviour[2], mechanical properties, liquid-phase[5,6] and scotch-tape exfoliation[7,8], and crystal growth.

Twin defects are identifiable by the presence of a twin boundary: a plane of lattice points where the crystals on either side of the interface possess mirror symmetry.[9] The formation of atomically sharp twin boundaries is ubiquitous in virtually all 3D materials including metals[10–12] and ceramics[13–15]. Our understanding of twinning in graphite has stemmed from the seminal work on twin structures by Friese and Kelly[4], nearly six decades ago, and has changed remarkably little since[9,16,17] although recently there has been significant interest in the electronic transport properties of wrinkles, bends and kinks in graphene[2].

2D materials are characterised by strong in-plane covalent bonding with weaker van der Waals bonding between the basal planes. The deformation behaviour of these crystals is therefore highly anisotropic. Slip occurs readily on the basal plane[18] but is forbidden on prismatic planes as this would require breaking the strong in-plane bonds[4,9]. In contrast, metallic bonding allows many crystallographic slip systems to be active simultaneously, leading to the ductile deformation behaviour of many metals. Twinning in metals occurs frequently to accommodate the strains induced during crystal growth and deformation, with atomic imaging of metallic twin boundaries[11,12] invariably revealing atomically-sharp crystal interfaces[11,12] illustrated in material science text books (Figure 1f inset blue box)[3,9]. Twinning in polymers is known to take place by the bending of the molecules across the twin boundary[19] but no one has previously performed an atomically-detailed analysis of twin boundaries in van der Waals crystals of 2D materials.

Bending of layered crystals requires neighbouring basal planes to slide past each other, usually as a result of the movement of basal dislocations. It is well known that 2D crystals with hexagonal unit cells, such as graphene, boron nitride and many transition metal dichalcogenides, have preferential low energy edges commonly termed armchair (ac) and zigzag (zz). Armchair and zigzag directions are preferred for basal plane dislocation movement and hence serve as preferential directions for folds, bends and kink bands when the material is stressed[20,21]. When basal plane dislocations stack vertically, a twin boundary can appear. This is defined crystallographically as s an interface separating two crystals with perfect commensurate stacking and mirror symmetry (see SI section 1). The formation of such boundaries is driven by energy minimisation between the strain energy of the distorted lattice and the vdW energy between adjacent planes (see Supplementary Information Section 3). To minimise the latter, crystals maintain perfect commensurate layer stacking away from the boundary causing a periodic constraint on the magnitude of the slip between each successive plane. The magnitude of the twin angle, $\theta_t$, for an ideal twin boundary is hence simply related to the quantised slip translation vector, $L_{slip}$, through the relationship

$$\theta_t = 2\, tan^{-1}\left[\frac{L_{slip}}{c}\right] \qquad (1)$$

where $c$ is the plane-normal lattice vector of the bulk crystal unit cell, equivalent to two basal plane separations for the 2D materials considered in this work (further details in Supplementary Information Section 1). The predicted values of $L_{slip}$ and $\theta_t$ for the zz and ac direction twin boundaries in graphite, hBN and $MoSe_2$ are listed in Table 1.

In this paper we have performed an extensive investigation of atomic morphology of deformation bends in exfoliated vdW materials and have found that twin boundaries are common microstructural features whose presence has previously been largely missed or ignored. In Fig. 1 we present atomic resolution cross sectional scanning transmission electron microscope (STEM) characterisation of kink bands in vdW crystals. Kink bands are deformation features that are often visible in optical microscope images of exfoliated flakes of 2D materials appearing as straight line striations ~1 μm wide (Fig. 1a). These appear along ac or zz directions and are formed by basal plane dislocations[22,23] due to applied thermal or mechanical stress. We have discovered that such defects predominantly consist of regions of perfect crystal lattice separated by boundaries where the basal planes are locally bent to the crystallographic twin angle (Equation 1 and Table 1). However, it is clear that, when imaged at the atomic scale, these twin boundaries are not abrupt classical twins but delocalised over tens of atoms; in the boundary region the perfect stacking of the neighbouring crystals is disrupted (Fig. 1e,f). Atomic resolution on both sides of the boundary can also be used to confirm commensurate crystal stacking (Fig 1e). However, this is often impossible to obtain in TEM or STEM images due to local strain effects. When this is not present a twin boundary must be identified from the quantisation of the slip vector ($L_{slip}$), measured from the difference in the length of neighbouring basal planes across the boundary (SI section 2).

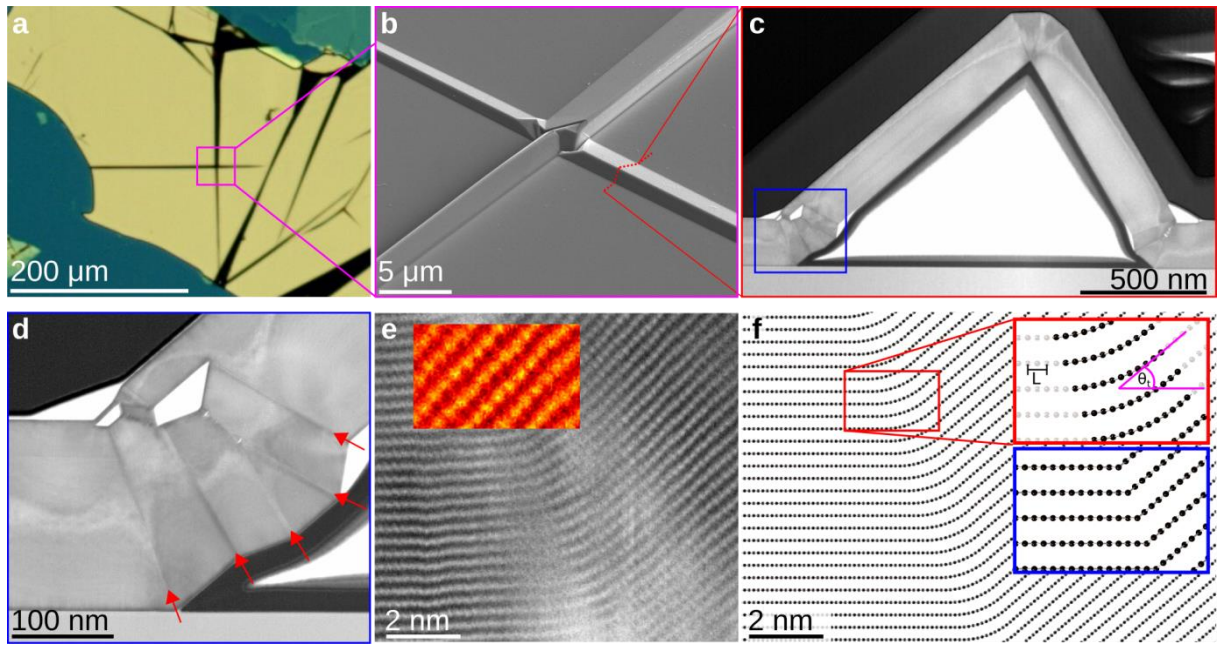

**Figure 1|Atomic structure of twin boundaries in layered vdW crystals. a**, Optical plan view image of a 200 nm thick graphite flake on silicon wafer. Stress induced kink bands can be seen intersecting one another at angular multiples of 30° (corresponding to zig zag, zz, or armchair, ac, crystal directions). **b**, Scanning electron microscopy (SEM) oblique view of zz kink band meeting an ac kink band at 90°. The site of the FIB cross section is annotated red. **c,d** Bright field STEM cross sectional imaging of the ac kink band. The kink is composed of flat areas of Bernal stacking bordered by many discrete boundaries of almost exactly the same angle, (boundaries annotated with red arrows) which are identified as incommensurate twin boundaries (see SI). **e** Filtered atomic resolution image of a twin boundary in hBN. The inset shows perfect atomic AA' stacking is maintained either side (parent and twin lattices) of the twin boundary. The core of the boundary is composed of a volume of crystal which bends like a nanotube and is incommensurately stacked. **f** Atomic model schematic comparing the delocalised incommensurate twin boundary structure (main figure and upper panel inset) with the conventional abrupt twin boundary (lower panel inset). The parameters that govern twin geometry, slip length $L$ and the ideal twin angle $\theta_t$ are annotated (see Table 1).

We have investigated the atomic morphology of bends, folds, and kink band ridges in graphene, hBN and MoSe$_2$. Performing lattice resolution STEM imaging of mechanically deformed crystals allows measurement of bend angles and slip translation vectors for individual boundaries. This analysis has revealed that all three vdW materials contain abundant twin boundaries despite their different stacking preferences, interlayer adhesion and bending moduli (Fig. 2). The ac twin boundaries shown in Fig. 2 have measured bend angles, θ, of ~40° ~39°, and ~30° for graphite, hBN and MoSe$_2$ respectively, in good agreement with theoretical predictions for $\theta_t$ of 40.3°, 41.2° and 30.0° (Table 1). Importantly, all twin boundaries demonstrate a curved region of incommensurate stacking with widths of 1.3 nm, 1.1 nm, 1.4 nm for graphite, hBN and MoSe$_2$ respectively. This delocalise interface is very different from conventional metallic twin boundaries, and perhaps explains why such features have not previously been identified as twin boundaries despite appearing in literature TEM images[21].

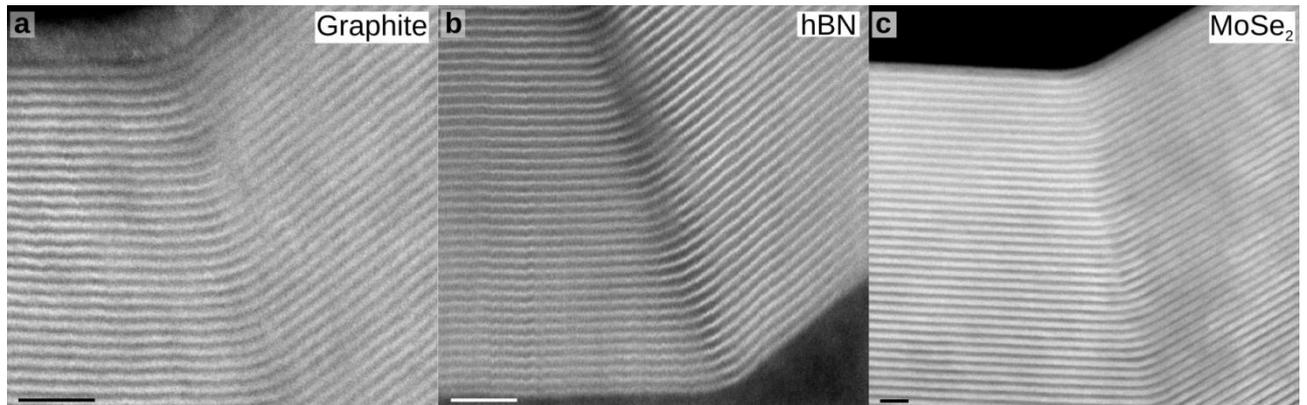

**Figure 2 | Cross sectional HAADF STEM images of delocalised twin boundaries in 2D materials. a**, in graphite; **b**, hBN; and **c**, MoSe$_2$. The viewing angle is down the armchair direction for all materials. This geometry is seen repeatedly in graphite kink bands as well as in van der Waals materials with different stacking preferences, interlayer adhesion and bending moduli. The bend angle in each material is dependent on its crystal lattice parameters (Table 1). All scale bars 2 nm.

|  | Graphite | | | | hBN | | MoSe$_2$ | |
|---|---|---|---|---|---|---|---|---|
| **Direction of twin boundary** | ac | zz | | | ac | zz | ac | zz |
| **Slip direction** | zz | ac | | | zz | ac | zz | ac |
| $L_{slip}$ | $a$ | $a/\sqrt{3}$ | $2a/\sqrt{3}$ | $3a/\sqrt{3}$ | $a$ | $3a/\sqrt{3}$ | $a$ | $3a/\sqrt{3}$ |
| $c$ (nm) | 0.67 | | | | 0.67 | | 1.23 | |
| $a$ (nm) | 0.25 | | | | 0.25 | | 0.33 | |
| $L_{slip}$ (nm) | 0.25 | 0.14 | 0.29 | 0.43 | 0.25 | 0.43 | 0.33 | 0.56 |
| $\theta_t$ (°) | 40.3 | 23.9 | 46.0 | 64.9 | 41.2 | 65.9 | 30.0 | 48.6 |

**Table 1**: Parameters calculated from literature X-ray crystallography structures for armchair and zigzag twin boundaries in graphite, hBN and MoSe$_2$. Higher order twin zz angles in graphite ($2a/\sqrt{3}$ and $3a/\sqrt{3}$) are likely to be less favourable due to the ease of splitting of such basal dislocations into separate partial dislocations with lower energy.

A more statistical analysis of a large number of single grain boundaries has revealed that bend angles cluster within ±18% of the ideal twin angle $\theta_t$ for all systems (for discussion see Supplementary Information Section 2). We propose that this variation in bend angle arises from the ability of the incommensurate region close to the twin boundary to accommodate significant local strain of the basal planes (for example monolayer graphene is known to accommodate high elastic strains up to 1.5%[24]) and the relatively low stacking fault energy of these materials.

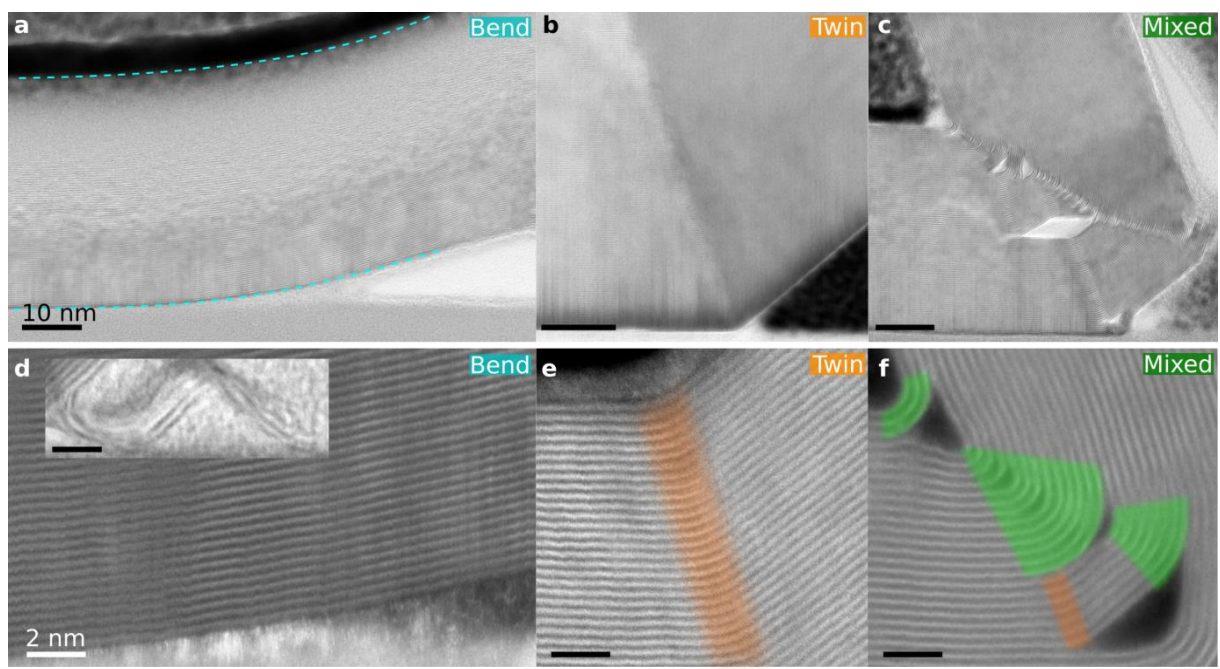

**Figure 3| Images of different bending phenomena in graphite. a-c** Bright field STEM images showing the microstructure of graphite for each bending mode: 'Bend', 'Twin' and 'Mixed' modes. The bend angle and number of basal planes in the crystal thickness for each are θ = 12°, $N_l$ =114 basal planes; θ = 40°, $N_l$ = 559; and θ = 115°, $N_l$ = 107 respectively. All scale bars 10 nm. **d-f** HAADF STEM images demonstrating the different ways van der Waals' crystals modify the arrangement of basal planes to accommodate strain induced by bending. The inset in **d** shows a 95° bend angle in a graphene bilayer. Areas of discrete twin boundaries are highlighted orange and areas of nanotube-like curvature are highlighted green. All scale bars 2 nm. The mixed bending mode exhibits multiple discrete twins converging on a region of nanotube-like curvature and is only observed for the largest bend angles.

Fig. 3 compares STEM images for sheets of graphite with different thicknesses subjected to increasing bend angle. At large bend angles we uncover an even more dramatic deviation from conventional twinning. Fig. 3f reveals the first observation of a mixed bending mode where basal planes in the lower part of the crystal form separate discrete twin boundaries (highlighted orange) converging to an extended region of uniform curvature towards the bend apex (highlighted green). In the latter, the material consists of concentric incommensurate basal planes with uniform curvature, akin to multi-walled nanotubes (see Supplementary Information Section 2).

Looking at a large number of the bending phenomena experimentally as a function of the number of atomic layers $N_l$ for the crystal flakes and their bend angle θ (Fig. 4) we can classify them as follows: **(a)** For all crystal thicknesses with small bend angles bending can be accommodated by strain in the lattice *(no twin boundary)*. **(b)** At small crystal thickness (few layers), nanotube-like bending is observed to accommodate bending of any angle, even greater than 90° *(no twin boundary)*. **(c)** For all but the thinnest flakes, *twin boundaries* are observed for angles close to $θ_t$, and **(d)** a mixed bending mode is observed for larger bend angles (*at least one twin as well as nanotube like bending*).

Using numerical modelling combined with first-principles calculations for stacking fault energies and bending moduli (SI Section 3), we can confirm our experimental observations. At small bend angles crystals of any thickness will remain commensurate by accommodating the shear strain through uniform nanotube-like bending of the basal planes, without using discrete local grain boundaries. For thicker flakes and bend angles approaching $θ_t$, it is necessary to consider the balance between the energy required to bend the basal planes, and that associated with creating regions of incommensurate stacking and/or stacking faults. For nanotube-like bending the total energy scales as $\sim(N_l)^2+\ln(N_l)$, while for a discrete twin boundary the total energy scales as $\sim N_l$. As a result nanotube-like curvature is more energetically favourable for thin crystals (small $N_l$) whereas discrete twinning becomes more favourable for thicker crystals (large $N_l$). The thickness threshold, $N_c$, at which nanotube-like curvature becomes unfavourable for graphite is predicted to be ~30 basal planes (blue dotted line in Fig. 4), in good agreement with experiment; the thinnest mixed-mode bending is observed in graphite flakes 31 basal planes thick.

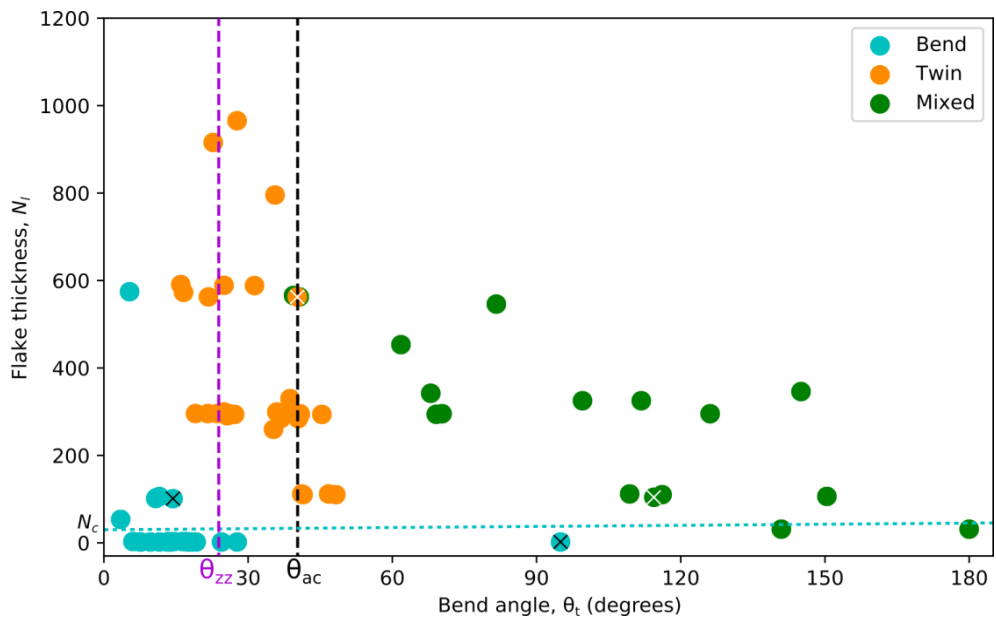

**Figure 4 | Plot of bending phenomena in graphite.** The plot demonstrates the distribution of 'Bend', 'Twin' and 'Mixed' modes of bending observed experimentally for different bend angles and thicknesses of crystal for both ac and zz bending. The dashed black and purple lines are the theoretically predicted ideal bend angles θ$_t$ for ac and zz twins respectively. The theoretical minimum thickness threshold for the twin or mixed modes is denoted by the dotted blue line. The data-points derived from the illustrative images in Fig. 3 are annotated with a cross. For flakes of any thickness, if θ < θ$_t$ the crystal can accommodate the strain with gentle bending whilst being fully commensurate, and for thin flakes, any θ > θ$_t$ results in nanotube-like bending of the basal planes (blue plot; 'Bend' mode). Above a flake thickness $N_c$ and around or below θ$_t$, discrete twin boundaries are found (orange plot; 'Twin' mode). Above θ$_t$ a third configuration is observed, the Mixed bending mode, for thick flakes (green plot; 'Mixed' mode).

For moderately thick flakes ($N_l > N_c$) at angles significantly greater than the twin angle we find that it is energetically favourable for the crystal to form multiple twin boundaries and/or exhibit the mixed bending mode shown in Fig. 3c,f. For mixed mode deformation the nanotube-like structure (incommensurate basal planes with high curvature) is only favourable for a limited number of atomic layers ($N_0$) closest to the bend apex. Further from the bend apex (*plane index*, $N > N_0$) the defect splits into discrete lower angle twins that locally are similar to those described in Fig. 2. Our theoretical calculations predict a transition thickness for graphite of $N_0 \approx 16$ in reasonable agreement with the experimentally observed transition at $N_0 \approx 10$ shown in Fig. 4 (see Supplementary Information Section 3). The difference in these values likely arises from our calculations, which cannot account for in-plane strain.[25] The transition between the nanotube-like curvature and discrete twin boundaries is analogous to the appearance of radial corrugations in large-diameter multi-walled nanotubes.[26]

To the best of our knowledge this is the first time such mixed mode bending has been observed and characterised. Small cracks are often seen at the interface between discrete twins and nanotube-like bending in all three materials. We speculate that the interface between nanotube-like bending and discrete twin regions may act as a weak point for nucleation of interlayer cracking; a process thought to be fundamental to exfoliation routes for 2D materials.[5–8] Crystals are also heavily damaged where twins along different crystallographic directions cross each other (such as at intersecting kink bands, Fig. 1b) which may lead to fragmentation of the crystals during exfoliation (see Section 5 in Supplementary Information). Furthermore, we observed that larger delamination cracks are formed at angles greater than $3\theta_t$ for crystals with thicknesses greater than $N_c$ (Supplementary Information Section 4). The absence of delamination cracks in thin crystals could help explain why liquid phase exfoliation of 2D materials struggles to yield large area flakes thinner than ~7 layers.[5]

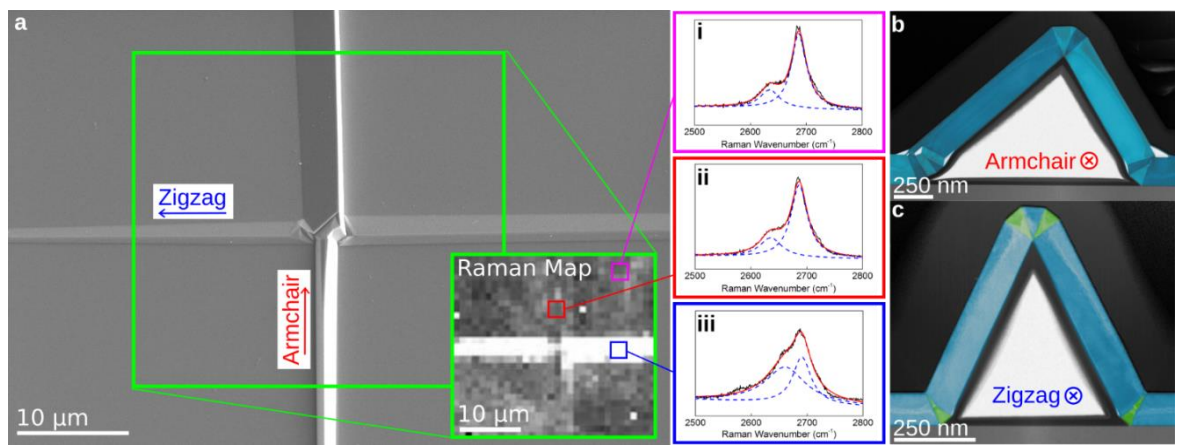

**Figure 5| a**, SEM image of two orthogonal kink bands intersecting in the crystal. Inset maps the Raman intensity ratio of the 2630 cm$^{-1}$ and 2690 cm$^{-1}$ peaks in the 2D peak. The ratio is drastically modified locally to the zz kink band (white contrast in inset). Individual Raman 2D spectra are shown for **i)** pristine graphite, **ii)** the ac twin and **iii)** the zz twin boundaries. **b**, Large field of view bright field STEM image of the ac kink band. In this striation the basal stacking in each crystalline region is equivalent (Bernal stacked). False colour is used to highlight areas of lattice separated by a twin boundary. This kink band shows some delamination yet gives the same 2D Raman signal as pristine graphite. **c**, Large field of view bright field STEM image of the zz kink band. Regions of Bernal stacking are false coloured blue while small areas of crystal that do not exhibit Bernal stacking and are instead highly faulted or AA' stacked are false coloured green. These areas are thought to contribute to the modification of the Raman 2D signal.

Finally we demonstrate that zz twin boundaries in graphite can induce local stacking faults and thus the Raman signature of zz kink bands can be used to determine crystallographic directions in graphite. Fig. 5a shows an SEM image and Raman imaging of zz and ac direction kink bands meeting at 90° in a thick flake of graphite. Analysis of the shape of the Raman 2D band reveals significant changes associated with the zz kink band but not for the ac direction kink band. The disruption of the Raman peak can be explained if we consider the formation of a twin boundary through a series of lattice translations along the favoured crystal directions. A sequence of three zz ideal twin boundaries (each with the expected $a/\sqrt{3}$ slip translations along the perpendicular ac direction) theoretically produces a region of crystal with a complex stacking sequence AABBCC (derived in Supplementary Information Section 3). This is a highly unfavourable stacking configuration in graphite. We therefore propose that it is more energetically favourable to induce stacking faults in small regions of the crystal between pairs of twin boundaries[27] (highlighted by the green shading on Fig. 5c). It is well established that a change in stacking in graphite has a strong effect upon the Raman signature, consistent with the changes we observe experimentally for the zz kink band (Fig. 5a and Supplementary Information Section 6).[22,28] This frustrated stacking will not be present in hBN or the transition metal dichalcogenides where AA' stacking is preferred[2] but we note that 2D materials with similar stacking preferences to graphite may exhibit comparable lattice disruption.

In conclusion, we have presented an atomic scale analysis of stress induced twin and mixed mode boundaries in van der Waals crystals of 2D materials, revealing a hitherto unsuspected level of unusual behaviour that we have been able to analyse and understand as a whole for the first time. We have discovered that the deformation microstructure depends on the mechanical properties of the material, the crystal thickness and bend angle, and deformation features can be predicted from numerical and first principles calculations. Thin crystals present nanotube-like bending while thicker crystals show discrete twins at low angles or mixed mode deformation at high angles. We find that a weak interface exists in mixed-mode bending that may act as a nucleation site for exfoliation. Our results provide fundamental insights for understanding the mechanisms of exfoliation for 2D crystal flakes, which are crucial to the development and application of 2D materials. Moreover, the crystallographic bending features and local stacking variations observed here have important implications for the electronic properties of the materials. For instance, graphene has been predicted to form p-n junctions[29] when bent and a similar effect in thicker films can partially account for observed modifications of the 2D Raman band.[29,30] It is also anticipated that transition metal dichalcogenides undergo significant change of their electronic spectrum on bending[31], which could be exploited for sensors or flexible electronics applications.

**Methods**

Thin graphite, hBN and MoSe$_2$ single crystal flakes were prepared on separate silicon substrates by mechanical exfoliation using adhesive tape.[16] For cross sectioning, the flakes were coated with 7 nm of amorphous carbon and 3 nm of Au/Pd to protect them during imaging and milling. A FEI Helios Nanolab DualBeam 660 instrument (incorporating a scanning electron microscope (SEM) and a focused ion beam (FIB) in the same chamber) was used to image the flakes and produce cross sections. Large, uniform striations seen in almost all thick flakes were identified by SEM imaging and with tilting of the SEM stage.

Cross sections were prepared perpendicular to the striations within the flakes. A 2 µm thick platinum layer was deposited over the region of interest (ROI) using a gas injection system and by patterning with the Ga$^+$ FIB. Large trenches are milled either side of the ROI to a depth of 5 µm using high current and high energy ions to leave a window comprising (ordered surface to substrate) platinum, Au/Pd, carbon, crystal striation, native silicon oxide and the silicon substrate. Once the FEI Easylift™ nanomanipulator was inserted and welded to the platinum layer with extra platinum, the window was cut free from the substrate and transferred to a pillar on a specialist OmniProbe™ TEM grid. The window was thinned to a TEM-transparent lamella with a thickness below 100 nm using 30 kV, 16 kV, 5 kV and 2 kV ion beam milling and polishing.

For high resolution scanning transmission electron microscope (STEM) imaging a probe side aberration corrected FEI Titan G2 80-200 kV was used with a probe convergence angle of 21 mrad, a HAADF inner angle of 48 mrad and a probe current of ≈80 pA. To ensure the electron probe was parallel to the basal planes, the cross-sectional sample was aligned to the relevant Kikuchi bands of the Si substrate and the 2D crystal.

The Raman spectra were obtained from the (0001) surfaces of the 2D crystals using a Renishaw inVia Raman system or Horiba LabRAM HR Evolution spectrometer, both equipped with HeNe lasers (λ = 633 nm), with the diameter of the laser spot on the crystal was estimated to be around 2 µm.

The density functional theory (DFT) calculations were performed using the periodic plane-wave basis set code VASP 5.35 and projector-augmented-wave potentials. The exchange–correlation functional is described by the revised Perdew–Burke–Ernzerhof (PBE) exchange model with the empirical dispersion correction of Grimme (DFT-D2). A plane-wave cut-off of 400 eV was used for the graphene and hBN layers and a value of 280 eV was used for the $MoSe_2$ layers. We used a Monkhorst–Pack $k$-point grid of 11×11×1 per unit, which ensured that the bond lengths and energies converged to within 0.001 nm and 1 meV, respectively.